\documentclass[twocolumn]{article}    

\usepackage{authblk}
\usepackage{mathptmx}     
\usepackage{graphicx}
\usepackage{amsmath}
\usepackage{mathtools}
\usepackage[numbers]{natbib}
\usepackage{color}
\definecolor{darkyellow}{RGB}{102,102,0}
\definecolor{orange}{RGB}{255,165,0}
\definecolor{purple}{RGB}{102,0,153}
\usepackage[top=1.5cm, bottom=1.5cm, left=1.5cm, right=1.5cm]{geometry}

\begin{document}

\title{Model for electron emission of high-Z radio-sensitizing nanoparticle irradiated by X-rays}

\author[1]{R.Casta \thanks{Electronic address: romain.casta@irsamc.ups-tlse.fr}} 
\author[1]{J.-P.Champeaux}
\author[1]{P.Cafarelli}
\author[1]{P.Moretto-Capelle}
\author[1]{M.Sence}

\affil[1]{Laboratoire Collisions Agr\'{e}gats R\'{e}activit\'{e}, IRSAMC, CNRS, UMR 5589, Universit\'{e} de Toulouse, UPS, F-31062 Toulouse, France.}

\date{June 17,2014}

\maketitle

\begin{abstract}
In this paper we develop a new model for the electron emission of high-Z nanoparticle irradiated by X-rays. This study is motivated by the recent advances about the nanoparticle enhancement of cancer treatment by radiotherapy. Our original approach combines a pure probabilistic analytical model for the photon trajectories inside the nanoparticle and subsequent electron cascade trajectories based here on a Monte-Carlo simulation provided by the Livermore model implemented in Geant4. To compare the nanoparticle and the plane surface electron emissions, we also develop our model for a plane surface. 
Our model highlights and explains the existence of a nanoparticle optimal radius corresponding to a maximum of nanoparticle electron emission. It allows us to study precisely the nanoparticle photon absorption and electron cascade production depth in the nanoparticle. 
\end{abstract}

\section{Introduction}

Electron emission by X-ray irradiated nanoparticle is an important subject of interest in the context of radiotherapy cancer treatment enhanced by nanoparticles of \mbox{high-Z} material such as gold. It has been shown that nanoparticles can improve  cancer radiotherapy enhancing significantly cancer cell destruction  \citep{chithrani_gold_2010,Herold} and cancer healing on mice \citep{hainfeld2004},  comparatively to classical radiotherapy treatments. Few theoretical approaches are available concerning this improvement, for example hyperthermia \citep{Kennedy,sharma_newer_2009}  or radical production \citep{Carter} have been studied. One of them explains this increase of cancerous cells death by the nanoparticles emission of electrons. These electrons, released by the interaction of X-rays with nanoparticles, are supposed to be more efficient in the destruction of cancerous cells than the electrons released in water (or tissue) by the interaction of the same X-rays \citep{Brun_Damage_DNA_LEE,Sanche,Butterworth}.
But this electron emission is not well known and consequently nanoparticle critical parameters such as size or composition are often chosen following others criteria than electron emission: biocompatibility \citep{Vujacic,Schaeublin}, commercial availabilities, production methods.
To fill this gap many studies have been done using Monte-Carlo simulation methods \citep{Casta,Chow,McMahon,Lechtman2013,Lechtman2011,Garnica} with particle transport codes like Liver\-more-Geant4 or PENELOPE-Geant4 \\ \citep{geant41,geant42,Penelope,PhysicG4}. These simulations are time consuming, strongly code and model dependent and do not allow the easy analysis of key parameters such size and composition. Therefore to complete these purely Monte-Carlo simulation studies, we propose a  semi-analytical original approach for the electron emission of high-Z nanoparticles. This approach combines a pure analytical model for photon trajectories and an electron cascade model partly based here on Monte-Carlo Livermore-Geant4 simulations, but which can be obtained by other models.\\

In the first part of this paper, we develop the electron emission model for a nanoparticle and for an infinite plane surface. In a second part dedicated to discussion, we compare our model to Livermore-Geant4 results. We also compare nanoparticle and infinite plane surface electron emission and we analyse the place where electrons are produced inside nanoparticle and plane surface. Finally we analyse the influence of the nanoparticle radius on nanoparticle electron emission with our model.   

\section{Nanoparticle electron emission}

We develop in this section the semi-analytical model for the electron emission of a spherical nanoparticle.

\subsection{Photon trajectory probability}

 \begin{figure}
   \centering
   \includegraphics[width=0.30\textwidth]{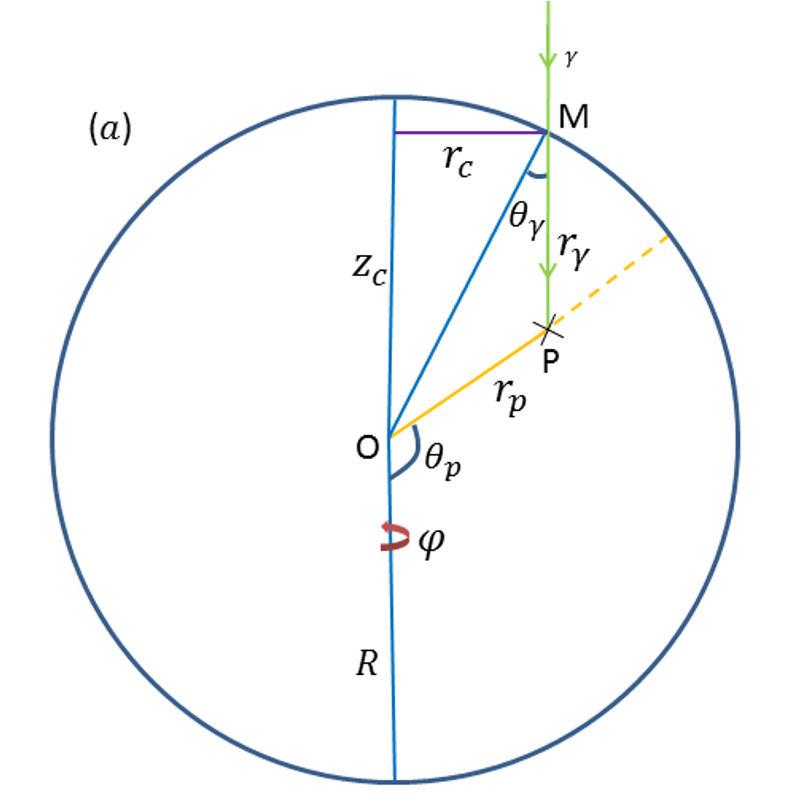}
    \includegraphics[width=0.22\textwidth]{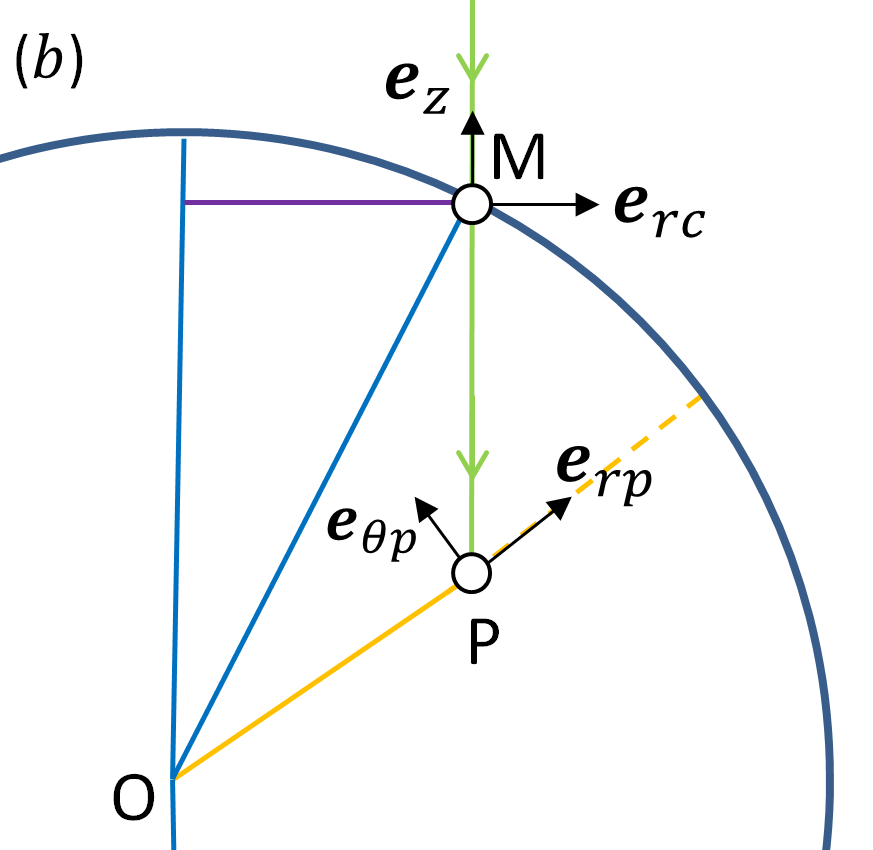}
    \includegraphics[width=0.30\textwidth]{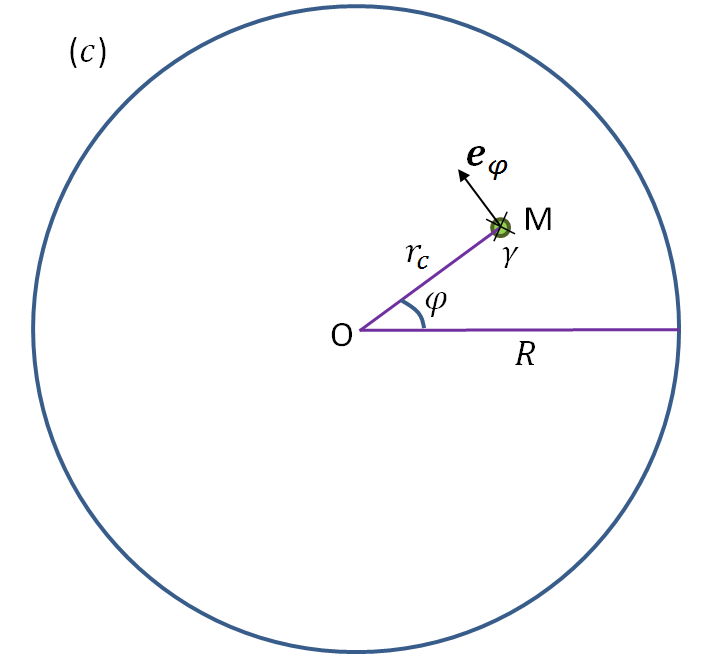}
     \caption{Photon trajectory in the plan $(MPO)$ through the nanoparticle $(a)$ and $(b)$ and the nanoparticle seen in the irradiated direction $(c)$.}
  \label{Fig:trajectorygeometry}
 \end{figure}

Our goal in this section is to determine the absorption probability density function (p.d.f.) of a photon in any point $P$ of the nanoparticle. We consider the trajectory of a photon of energy $E_\gamma$ incident to the nanoparticle at an entry point called $M$. The photon trajectory geometry is described on Fig.\ref{Fig:trajectorygeometry}. Subfigures $(a)$ and $(b)$ represent the $(MPO)$ plan  where $O$ is the nanoparticle center, $M$ is the entry point specified by the cylindrical coordinates system $(r_c,\varphi,z_c)$ of origin $O$ and local orthogonal unit vectors $(\vec{e}_{rc},\vec{e}_\varphi,\vec{e}_z)$. $P$ is the absorption point specified by the coordinates $(r_c,\varphi,z_c-r_\gamma)$ in the  cylindrical coordinates system or by the spherical coordinates system $(r_p,\theta_p,\varphi)$ of origin $O$ and local orthogonal unit vectors $(\vec{e}_{rp},\vec{e}_{\theta p},\vec{e}_\varphi)$. We have also represented on on Fig.\ref{Fig:trajectorygeometry} the radius $R$ of the nanoparticle and its surface by a blue circle.
The subfigure $(c)$ represents the nanoparticle seen in the irradiation direction i.e.  $\vec{e}_z$ direction. 
Considering that the irradiation source is far from the nanoparticle comparatively to its radius $R$, a very good approximation is to consider parallel photon irradiation.

\subsubsection{Incident and azimuthal angle probability density function (p.d.f.)}

If the photon trajectories are parallel and if the photons are uniformly distributed, we can write the probability $p(M)dS_M$ that a photon goes through an infinitesimal horizontal surface $dS_M= r_c dr_c d\varphi$ around a point $M$ as:

\begin{equation}
p(M)dS_M=\frac{dS_M}{\pi R^2}=\frac{r_c dr_c d\varphi}{\pi R^2}
\end{equation}

Writing $r_c$ as a function of $\theta_\gamma$: $r_c=R \sin(\theta_\gamma)$ (see Fig.\ref{Fig:trajectorygeometry}$(a)$) we then have:

\begin{align}
&p(r_c,\varphi) r_c dr_c d \varphi=p(\theta_\gamma,\varphi)d\theta_\gamma d \varphi \nonumber \\
&=\frac{r_c}{\pi R^2}\frac{\partial r_c}{\partial \theta_\gamma}d\theta_\gamma d\varphi= \frac{\sin (2\theta_\gamma)}{2 \pi} d\theta_\gamma d\varphi
\label{eq:angle}
\end{align}
which is the incident and azimuthal angle p.d.f..\\

\subsubsection{Photon path length $r_\gamma$ probability density function (p.d.f.)}

In this section we determine the p.d.f. that a photon goes through the nanoparticle on a distance $r_\gamma$ (see Fig.\ref{Fig:trajectorygeometry}$(a)$) and is absorbed.\\

Considering the X-ray photon energy and the nanoparticle high-Z uniform composition, we choose to consider only the photo-electric interaction for photons. We assume that photons are not scattered and nor they loose their energies on their tracks except if they are completely absorbed and stopped by a photo-electric process, and emit  electrons.

Following these approximations, we can write the expression of the probability that a photon goes through a distance $r_\gamma$ inside the high-Z material composing the nanoparticle and reacts as:

\begin{equation}
p(r_\gamma)dr_\gamma=\frac{\exp(-r_\gamma/\lambda_\gamma)}{\lambda_\gamma}dr_\gamma
\label{eq:rg}
\end{equation}
where $\lambda_\gamma$ is the photon mean free path inside the nanoparticle material.

The probability that the photon goes through a path length $r_\gamma$ is approximated by a classical Beer-Lambert law and the absorption probability is just $1/\lambda_\gamma$. The photon mean free path $\lambda_\gamma$ is strongly photon energy and nanoparticle material dependent.

\subsubsection{Absorption probability density function (p.d.f.)}

We study the probability that a photon be absorbed in a volume $dV_p$ around a point $P$ of the nanoparticle. Using (\ref{eq:angle}) and (\ref{eq:rg}) we can write the absorption p.d.f. in the cylindrical coordinates system:

\begin{align}
p(P) d V_p&=p(r_c,\varphi,z_c-r_\gamma)r_c dr_c d\varphi d(z_c-r_\gamma)  \nonumber \\
&=p(r_\gamma) p(r_c,\varphi)r_c dr_c d\varphi dr_\gamma \nonumber \\
&=\frac{\exp(-r_\gamma/\lambda_\gamma) \sin(2 \theta_\gamma)}{2 \pi \lambda_\gamma} d r_\gamma d\theta_\gamma d\varphi
\end{align}

We can also express the same $p(P)dV_p$ absorption  probability in the $(r_p,\theta_p,\varphi)$ spherical coordinate system:

\begin{align}
p(P) d V_p&=p(r_p,\theta_p,\varphi)r_p^2 \sin \theta_p dr_p d\theta_p d\varphi 
\end{align}

Finally by doing the change of variables $(r_\gamma,\theta_\gamma) \longrightarrow (r_p,\theta_p)$, we can express the absorption probability, as a function of $(r_p,\theta_p,\varphi)$, in an infinitesimal volume $dV_p$ around a point $P$ of the nanoparticle.:

\begin{equation}
\begin{split}
&p(r_p,\theta_p,\varphi)dV_p=p(r_p,\theta_p,\varphi)r_p^2 \sin \theta_p d r_p d \theta_p d \varphi \\
&=\frac{\exp(-r_\gamma/\lambda_\gamma)\sin(2 \theta_\gamma)}{2\pi \lambda_\gamma}\left|\frac{\partial(r_\gamma,\theta_\gamma)}{\partial(r_p,\theta_p)}\right|d r_p d \theta_p d \varphi 
\end{split}
\label{eq:abs}
\end{equation}

To complete this equation, we need to write $r_\gamma$ and $\theta_\gamma$ as a function of $r_p$ and $\theta_p$:
\begin{align}
&\theta_\gamma=\sin^{-1} \left( \frac{r_p \sin \theta_p}{R} \right)\\
&r_\gamma=\frac{R\sin (\theta_p+\theta_\gamma)}{\sin \theta_p}
\end{align} 

Leading to the corresponding Jacobian matrix determinant which is:
\begin{equation}
\left|\frac{\partial(r_\gamma,\theta_\gamma)}{\partial(r_p,\theta_p)}\right|=\frac{r_p}{R \cos \theta_\gamma}
\end{equation}

\subsubsection{Application: Absorption probability density function (p.d.f.) as a function of $r_p/R$}

\begin{figure}
   \centering
   \includegraphics[width=0.4\textwidth]{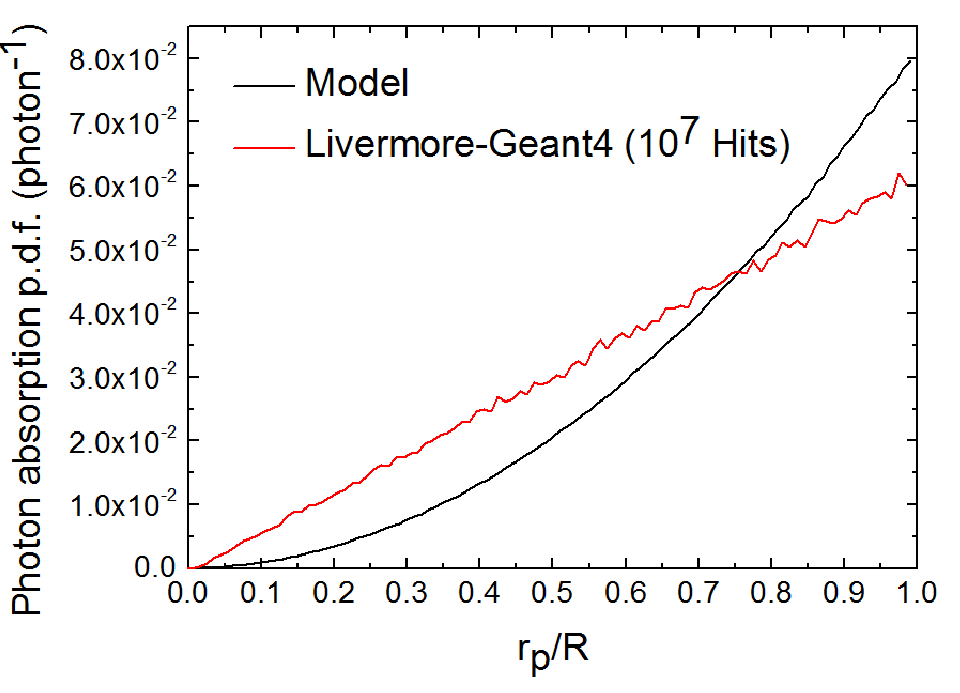}
     \caption{Absorption p.d.f. as a function of $r_p/R$ from our model (black) and from Livermore-Geant4 (red) for a gold nanoparticle of radius $R=5nm$ and $E_\gamma=1486.5eV$.}
  \label{Fig:Reaction_d.f.p.}
 \end{figure}

It is interesting to integrate the equation (\ref{eq:abs}) over $\theta_p$ in order to draw the photon absorption p.d.f. as a function of the relative distance $r_p/R$ to the nanoparticle center. We performed this integration with a Monte Carlo numerical method, as an example, for $R=5nm$, $E_\gamma=1486.5 eV$ and $\lambda_\gamma=247.874nm$ (corresponding to gold material) and we draw the result on Fig.\ref{Fig:Reaction_d.f.p.}.

To check our model results, we compare them with a Monte-Carlo particles transport simulation performed ten millions times with the Livermore Model implemented in Geant4 for the same $R$, $E_\gamma$, $\lambda_\gamma$ parameters. The Livermore model (Salvat et al, 2011; Wright, 2012) includes among others photoelectric process, electrons scattering and electron impact ionization processes. In our case, this model is implemented in the transport toolkit Geant4 (Agostinelli et al, 2003; Allison et al, 2006) that we use in its 4.9.6 version.\\

The $E_\gamma$ photon energy corresponds to the energy of a Al $K_\alpha$ X-ray source used in a previous work \citep{Casta} where we compared experimental gold nanoparticle and gold plane surface electron emission, with Monte-Carlo simulations.

We observe that the p.d.f. integral over $r_p/R$ is not equal to $1$. This is due to the small photon number absorbed (i.e. the integral over all $r_p/R$ values) by the nanoparticle, $2.6\%$ in our case.
The two curves show a difference between our model and the \mbox{Livermore-Geant4} simulation. Indeed for this last one, the p.d.f. is linear as a function of $r_p/R$ whereas the model one follows a quadratic function. This difference is not explained for the moment but originates probably in the Geant4 nanoparticle geometry management.
Nevertheless the two p.d.f. are very close which confirms our approach. The Livermore-Geant4 p.d.f. can be seen as a good approximation of the model one. \\

\subsection{Electron cascade trajectory probability density function (p.d.f.)}
 
 At the $keV$ energy range of interest, photon trajectories are quite simple: photons do not loose energy and are not scattered during their tracks. But their absorptions inside the nanoparticle give rise to the creation of free electrons by photo-electric processes. These primary electrons scatter inside the matter and create secondary electrons of lower energies by electron impact ionisation which in turn are scattered inside the matter and can themselves  create secondary electrons. Lets call this process an electron cascade. Some of these electrons can reach the nanoparticle surface and contribute to the nanoparticle electron emission.\\
 
\subsubsection{Electron cascade probability density function (p.d.f.) approximation from Livermore-Geant4 in the bulk solid} 

In this section we define and study the electron cascade p.d.f. in an infinite solid before including it in our nanoparticle model.

For a single absorption event, lets define at any point within the solid the electron cascade probability as the probability that an electron of energy $E$ produced in the cascade goes through the infinitesimal surface around this point. The electron cascade probability is given by:
 \begin{equation}
  \psi(E_\gamma,E,r,\Omega)d \Omega dE =\psi(E_\gamma,E,r,\theta,\varphi)\sin \theta d\theta d \varphi dE
   \label{eq:PS} 
 \end{equation}
 
\begin{figure}
   \centering
   \includegraphics[width=0.4\textwidth]{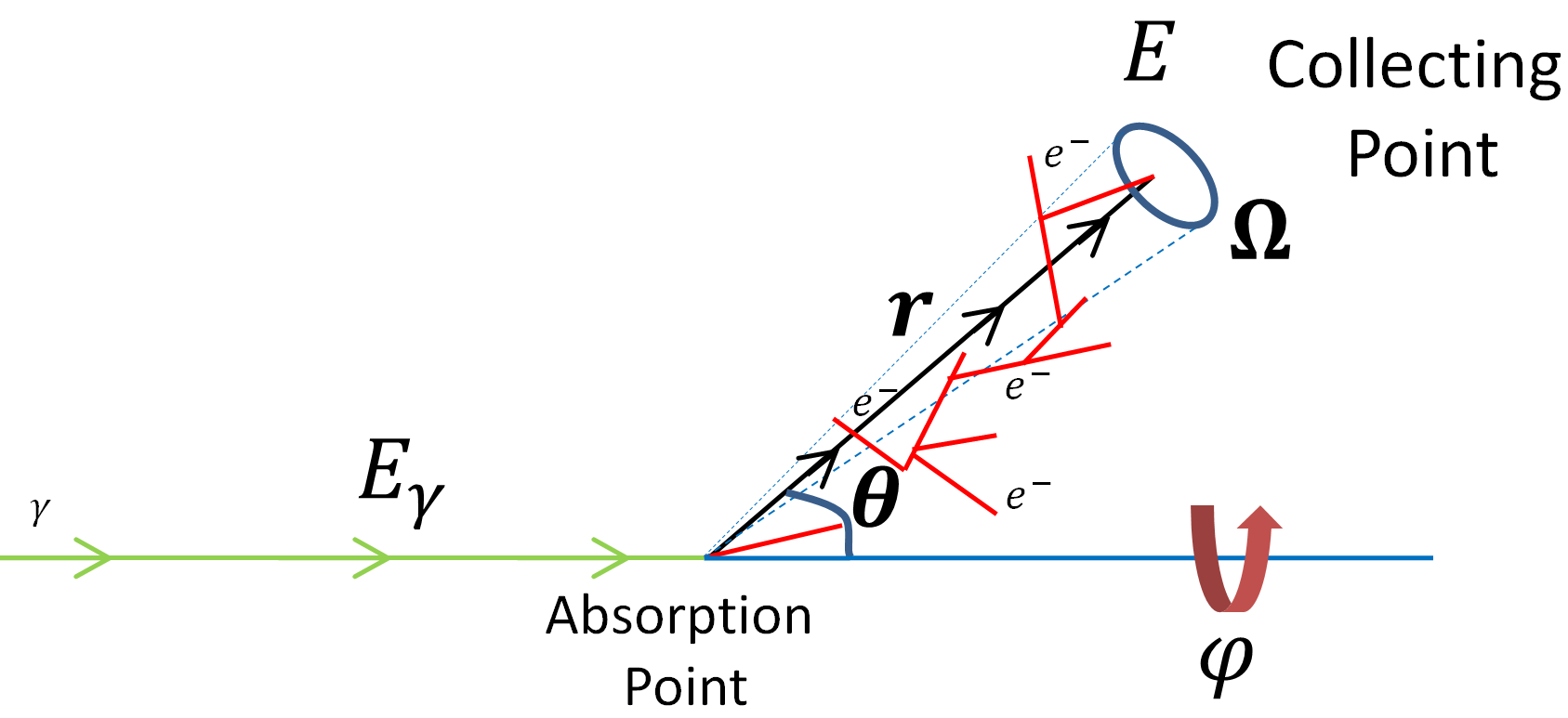}
     \caption{Electron cascade p.d.f. $\psi(E_\gamma,E,r,\theta,\varphi)$ schematic geometry. }
  \label{Fig:PsiGeom}
 \end{figure}

It is the probability to find an electron of energy $E$ in an infinitesimal solid angle $d\Omega$ after a photo-electric reaction of a photon of energy $E_\gamma$, given the distance $r$ to the absorption point, and the photon energy $E_\gamma$ as described on Fig.\ref{Fig:PsiGeom}.
 
Here we do not analytically compute the p.d.f. $\psi$ but we approximate it from Livermore-Geant4 by simulating the experiment shown on Fig.\ref{Fig:PsiGeom} fifty million times for each $r$ value with a low cutoff at 100eV for electrons and photons. We compute energy $E_e$, path length $r$ and scattering angle $\theta$ for all the electrons emitted forward i.e. the backscattered electrons are not scored and we assume that the $\varphi$ p.d.f. follows a uniform law. From these collected data we draw a $3D$ histogram. We choose relatively small bin sizes: $\Delta r=0.1nm$, $\Delta \theta=\pi/100$ and $\Delta E=1eV$, in order to have a good approximation. Each value of the p.d.f. $\psi$ is approximated to its corresponding bin value in the $3D$ histogram. The relative error is evaluated to a few percents by measuring the $3D$-histogram noise intensity. The Livermore-Geant4 model reliability has been successfully confronted to the experiment (see \cite{Casta}).\\

\begin{figure}
   \centering
   \includegraphics[width=0.4\textwidth]{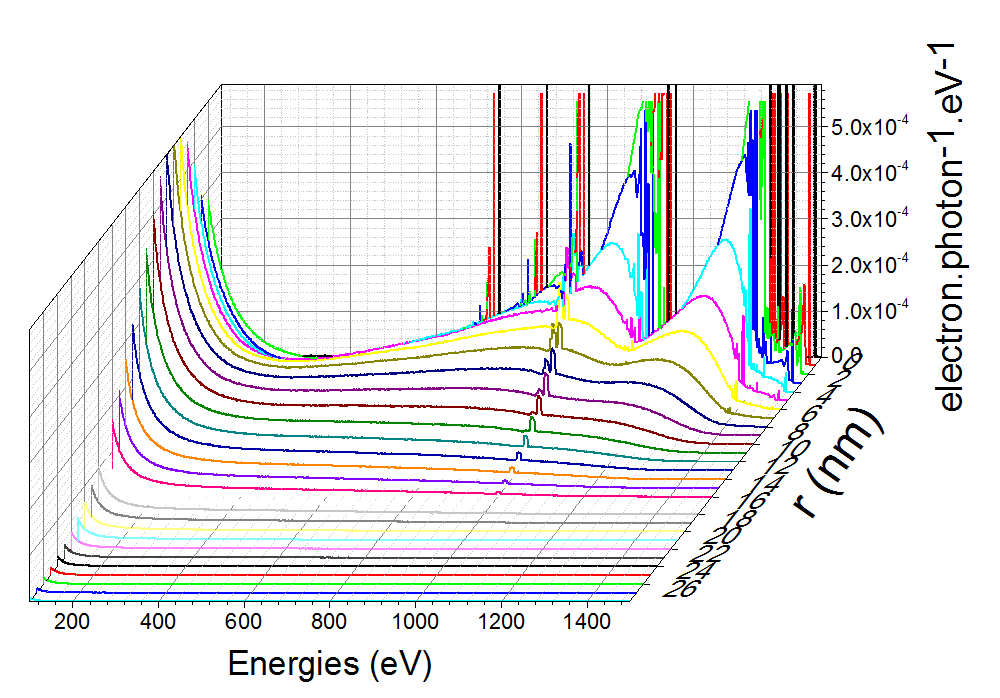}
     \caption{Electron cascade p.d.f. in gold for $E_\gamma =1486.7eV$ integrated over angles i.e. $\int \psi (E_\gamma ,E,r, \Omega) d \Omega $ for \mbox{$1nm \leq r \leq 28nm$} and \mbox{$150eV \leq E \leq 1500eV $} . }
  \label{Fig:PsiCascade}
 \end{figure}

Fig.\ref{Fig:PsiCascade} shows a few values of the approximated electron cascade p.d.f. in gold, integrated over angles $\theta$ and $\varphi$ for $E_\gamma=1486.5eV$. We observe that p.d.f. values are very small for $r>28nm$ compared to smaller $r$ values.  Therefore we approximate this p.d.f. to zero for $r$ values superior to $30nm$.

On this figure, $\int \psi d\Omega$ p.d.f. shows photo-electric peaks for the small $r$ values. They go broader as we move away from the absorption and finally merge into the electron background. These peaks are analysed in details in \cite{Casta}.

\subsubsection{Inclusion of the electron cascade probability density function (p.d.f.) approximation into the nanoparticle model}
 
Now that we have computed the electron cascade p.d.f. lets inject it in the nanoparticle model.
 
\begin{figure}
   \centering
   \includegraphics[width=0.4\textwidth]{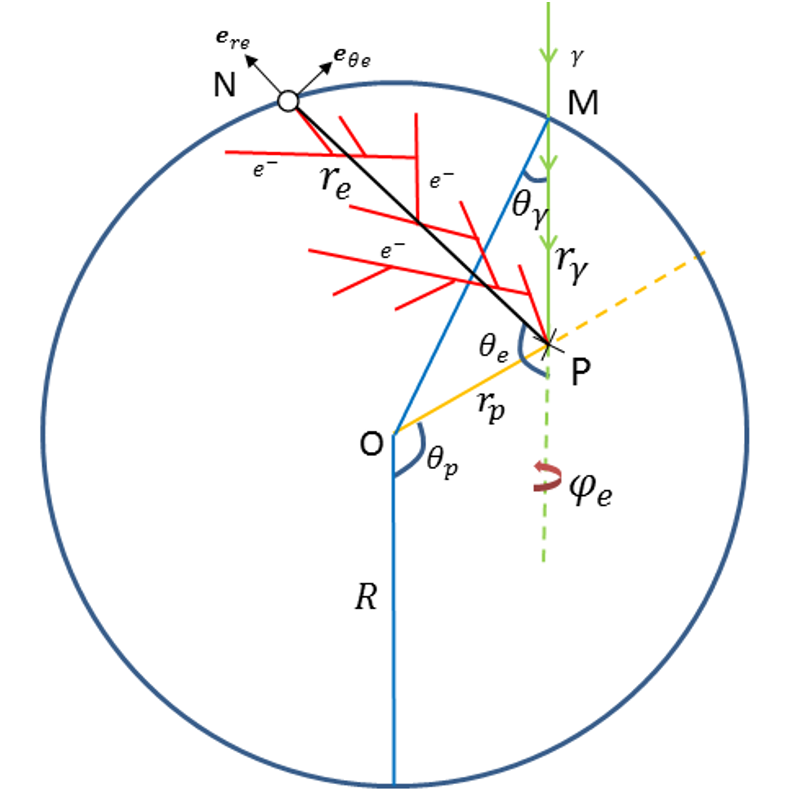}
     \caption{Photon and electron cascade trajectories in a particular cross-section.}
  \label{Fig:TotalGeom}
 \end{figure}

 We represent the photon and electron cascade trajectories on Fig.\ref{Fig:TotalGeom} where $r_e$ is the distance from the absorption point $P$ to a nanoparticle surface point $N$ specified in the spherical coordinates system $(r_e ,\theta_e, \varphi_e)$ of origin $P$ and local orthogonal unit vectors $(\vec{e}_{re},\vec{e}_{\theta e},\vec{e}_{\varphi e})$. 
 
 We can associate the electron cascade p.d.f. defined in (\ref{eq:PS}) to the electron cascade trajectory $PN$. We write the probability to find an electron of energy $E_e$ in the infinitesimal solid angle $d \Omega_e=\sin \theta_e d\theta_e d\varphi_e$ around $N(r_e,\theta_e,\varphi_e)$ given the length $r_e$ between the photon absorption point $P$ and the surface point $N$, and the photon energy $E_\gamma$:
 
\begin{equation}
 \psi(E_\gamma,E_e,r_e,\theta_e,\varphi_e)\sin \theta_e d\theta_e d \varphi_e dE_e
 \label{eq:Psi}
 \end{equation} 

 \subsection{Electron emission probability density function (p.d.f.)}
 
 We can multiply photon absorption and electron cascade p.d.f. in order to obtain the  probability of an incident photon entering in the nanoparticle at $M$, absorbed at $P$ and producing an electron cascade ejecting an electron of energy $E_e$ at a surface point $N$:
 
 \begin{equation}
 \begin{split}
 &\frac{e^{-r_\gamma/\lambda_\gamma}\sin(2 \theta_\gamma)}{2\pi \lambda_\gamma}\left|\frac{\partial(r_\gamma,\theta_\gamma)}{\partial(r_p,\theta_p)}\right|d r_p d \theta_p d \varphi\\
 &\times \psi(E_\gamma,E_e,r_e,\theta_e,\varphi_e)\sin \theta_e d\theta_e d \varphi_e dE_e
 \end{split}
 \label{eq:tot}
 \end{equation}

$r_e$ can be written as a function of the integration variables:
\begin{equation}
\begin{split}
r_e=&-r_p (\cos \varphi_e \sin \theta_e \sin \theta_p+ \cos \theta_e \cos \theta_p )\\
&+\sqrt{r_p^2 \left( 
\begin{split}
&\cos ^2 \varphi_e \sin ^2\theta_e \sin^2\theta_p\\
&+0.5 \cos \varphi_e  \sin (2\theta_e)  \sin (2\theta_p) \\
&+\cos ^2 \theta_e \cos^2 \theta_p -1
\end{split}
\right)+R^2}
\end{split}
\end{equation} 
 
We can integrate (\ref{eq:tot}) over the variables ($r_p$, $\theta_p$, $\varphi$) and ($\theta_e$,$\varphi_e$) corresponding respectively to the absorption points $P$ and the surface points $N$. Hence we get the total nanoparticle electron emission at an energy $E_e$:
 
 \begin{align}
&p(E_e)dE_e= \left( \int_P \int_N \frac{e^{-r_\gamma/\lambda_\gamma}\sin(2 \theta_\gamma)}{2\pi\lambda_\gamma}  \left|\frac{\partial(r_\gamma,\theta_\gamma)}{\partial(r_p,\theta_p)}\right| \right. \nonumber \\
&  \times \psi(E_\gamma,E_e,r_e,\theta_e,\varphi_e)\sin \theta_e d\theta_e d \varphi_e d \theta_p  d \varphi d r_p \Bigg) dE_e 
 \end{align} 
 
By replacing $\psi(E_\gamma,E_e,r_e,\theta_e,\varphi_e)$ by its Livemore-Geant4 approximation, we can numerically compute this integral with a Monte-Carlo method. We do it on a large range of energies and we draw the resulting electron emission intensity as a function of the electron energies.

\begin{figure}
   \centering
   \includegraphics[width=0.4\textwidth]{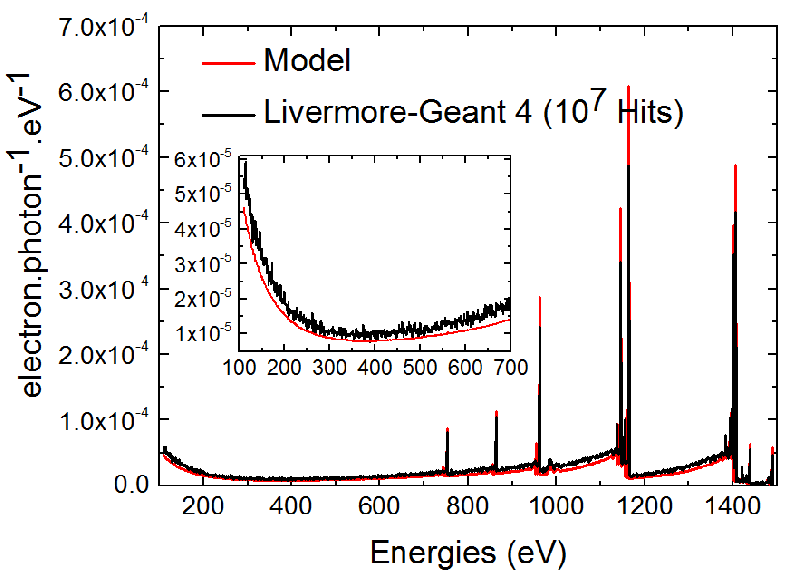}
     \caption{Model and Livemore-Geant4 electron emission spectra for a gold nanoparticle, $R=5nm$ and $E_\gamma=1486.5eV$. The spectra are zoomed to the range $100eV-700eV$ in the upper box.}
  \label{Fig:spectrum}
 \end{figure}

The obtained spectrum is shown on Fig.\ref{Fig:spectrum} for a X-ray energy value of  $E_\gamma=1486.5eV$ and a gold nanoparticle of $R=5nm$ as an example. It is compared with a purely Livermore-Geant4 simulation of ten millions hits, with the same parameters. The computing times comparison is clearly at the advantage of the model. For the same relative standard deviation, it takes $38min$ to compute the whole electron emission spectrum with the Livermore-Geant4 model whereas it takes only $90s$ with our model. This gain in time can be largely improved by refining the integration method. The computations were performed on a computer equipped with an Intel Core i7-3770 processor and 16GB memory.

We can see that  both spectra are very close but present a few differences which are explained by the photon absorption p.d.f. (Fig.\ref{Fig:Reaction_d.f.p.}) in a following section.\\

Our model is reliable as long as the photo-electric process is the main photo-reaction process of photons inside the nanoparticle material. The electron trajectory reliability is fully dependent of the model used. In the Livermore-Geant4 case, we assume that the electron trajectory simulations are correct until $100eV$. In our specific case, the model is valid in the photon energy range $100eV-1MeV$, and for electron of energies higher than $100eV$.\\
This work can be achieved for higher photon energy used in radiotherapy treatment with the same method and with the Livermore-Geant4 model for example. This will be the subject of a next paper.

\section{Plane surface electron emission}

In a previous paper \citep{Casta}, we compared nanoparticle of radius $19nm$ and very large gold plane surface experimental electron emission spectra. To complete this work, we have decided to develop this model for plane surface electron emission too.

\begin{figure}
   \centering
   \includegraphics[width=0.4\textwidth]{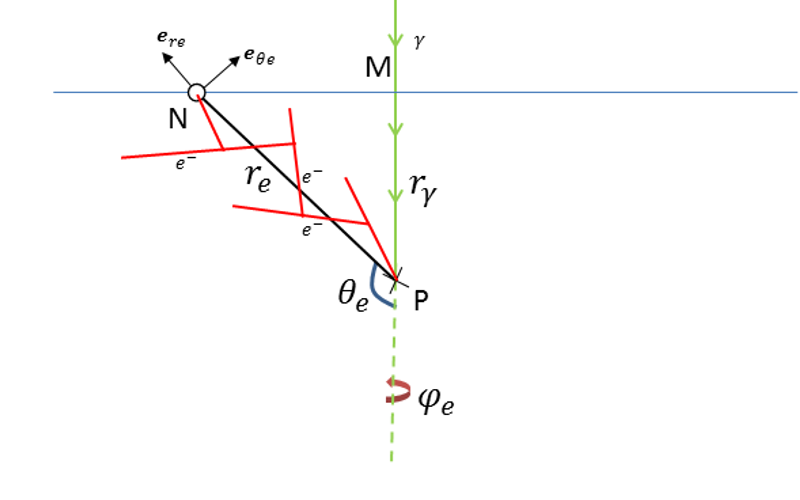}
     \caption{Infinite plane surface photon and electron trajectories.}
  \label{Fig:cuboid_geom}
 \end{figure}

 As well as for nanoparticles we consider the photon and the electron cascade trajectories. These are represented on Fig.\ref{Fig:cuboid_geom}. $M$ is the entry point, $P$ the absorption point specified by its depth $r_\gamma$ and $N$ the exit point specified by the spherical coordinates system $(r_e,\theta_e,\varphi_e)$ of origin $P$ and local orthogonal unit vectors $(\vec{e}_{re},\vec{e}_{\theta e},\vec{e}_{\varphi e})$.

In the same approximations than for nanoparticles and by integrating over all the surface points $N(\theta_e,\varphi_e)$ and the depth $r_\gamma$, we express the total infinite plane surface electron emission p.d.f. for an energy $E_e$ as:

 \begin{align}
&p(E_e)dE_e= \Bigg( \int_{\pi/2}^{\pi} \int_0^{\infty} \frac{e^{-r_\gamma/\lambda_\gamma}}{\lambda_\gamma} \nonumber \\
& \times 2\pi \psi(E_\gamma,E_e,r_e,\theta_e,\varphi_e)\sin \theta_e dr_\gamma d\theta_e \Bigg)  dE_e
\end{align}

Again we have $r_e$ as a function of $r_\gamma$ and $\theta_e$:
\begin{equation}
r_e=-\frac{r_\gamma}{\cos{\theta_e}}
\end{equation}
 
\begin{figure}
   \centering
   \includegraphics[width=0.4\textwidth]{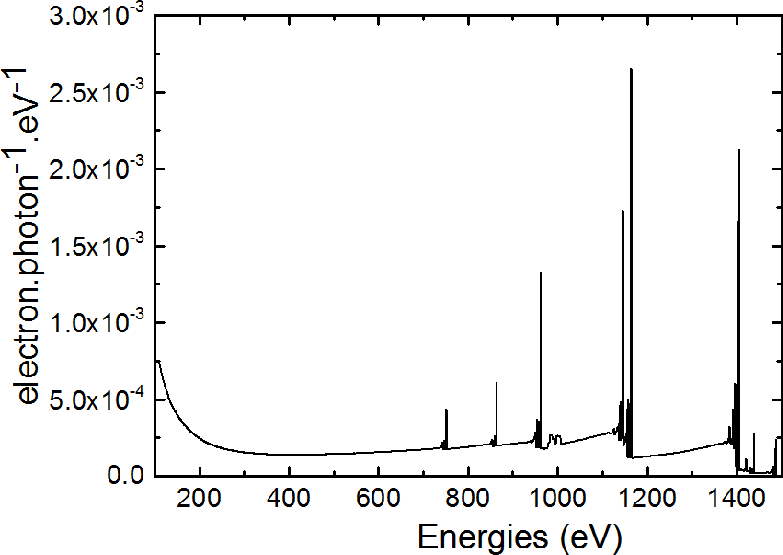}
     \caption{Electron emission spectrum for a gold plane surface.}
  \label{Fig:Plane_spectrum}
 \end{figure}

We integrate this p.d.f. by a Monte-Carlo numerical method for a gold plane surface as an example. The result is shown on Fig.\ref{Fig:Plane_spectrum}.
 
 We can see that the electron emission for the plane surface is much larger than for the $5nm$ radius nanoparticle (Fig.\ref{Fig:spectrum}). For example the surface electron emissions are $1.38 \times 10^{-4}$ $electron.photon^{-1}.eV^{-1}$ at $400eV$ and $2.65 \times 10^{-3}$\\ $electron.photon^{-1}.eV^{-1}$ at $1164eV$ whereas for the nanoparticle the electron emissions are respectively $8.07 \times 10^{-6}$ and $6.09 \times 10^{-4}$. This is explained by the larger photon number absorbed by the gold surface plane than by the gold nanoparticle.
 
Indeed, because of its infinite depth all the incident photons are absorbed by the plane surface, producing photo-electrons whereas only $2.6 \%$ of the incident photons are absorbed by the nanoparticle of radius $5nm$. We do a more extensive comparison in a further section.  
 
  \section{Discussion}

\subsection{Nanoparticle electron cascade production depth} 
 
\begin{figure}
   \centering
   \includegraphics[width=0.4\textwidth]{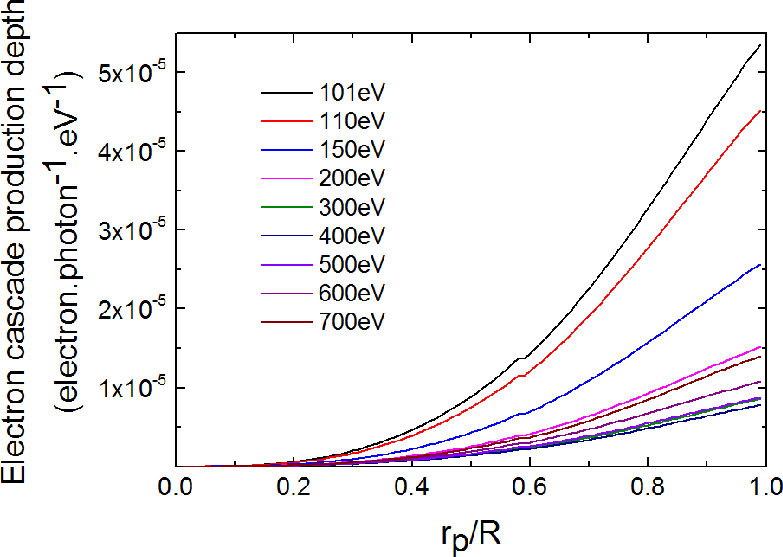}
     \caption{Electron cascade production depth p.d.f. i.e. p.d.f. that a photon of energy $E_\gamma$ reacts at a distance $r_p$ from the gold nanoparticle center and produces an electron of energy $E_e$ at the nanoparticle surface as a function of $r_p/R$, for $R=5nm$ and $E_\gamma=1486.5eV$.}
  \label{Fig:dpf_r}
 \end{figure}

Before going further, we choose to study the electron cascade production depth p.d.f..
To obtain this p.d.f., we numerically integrate equation (\ref{eq:tot}) on all the variables excepted $r_p$ and $E_e$ and draw the resulting p.d.f. which represents the electron cascade production depth i.e. the probability that a photon of energy $E_\gamma$ reacts at a distance $r_p$ from the nanoparticle center and produces an electron of energy $E_e$ at the nanoparticle surface. We draw it on Fig.\ref{Fig:dpf_r} for a gold nanoparticle  of radius $5nm$ irradiated by $1486.5eV$ photons, and few electrons energies located in the electron background range $100eV-700eV$.
 
These electron cascade production depth p.d.f. show, as expected, that most of the emitted electrons are produced by absorption close from the nanoparticle surface. After normalization we notice that all of them follow the same non-linear function which has a maximum at the nanoparticle surface.\\

\begin{figure}
   \centering
   \includegraphics[width=0.4\textwidth]{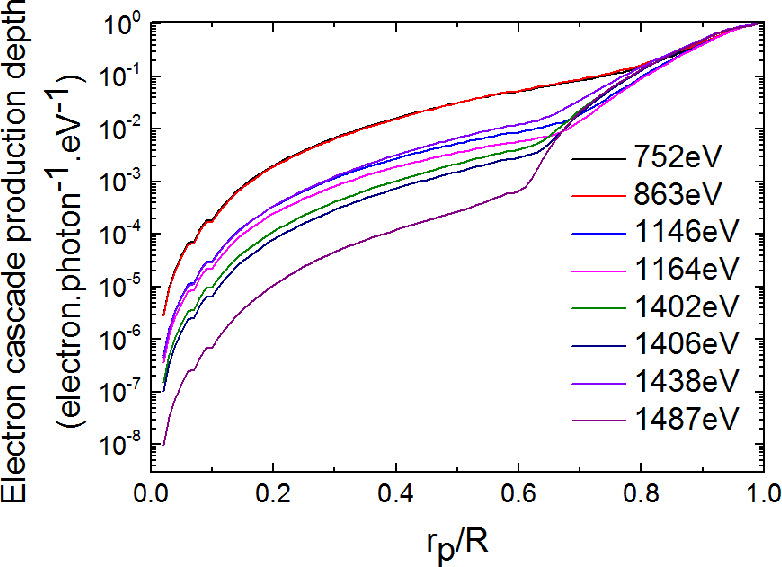}
     \caption{P.d.f. that a photon of energy $E_\gamma=1486.5eV$ reacts at a distance $r_p$, from the center of a gold nanoparticle of radius $R=5nm$, and produces an electron of energy $E_e$ at the nanoparticle surface as a function of $r_p/R$ for photo-electric lines.}
  \label{Fig:dpf_r_peak}
 \end{figure}

This non-linear function is not completely described by the electron cascade production depth p.d.f. of the photo-electric peak energies as we see on Fig.\ref{Fig:dpf_r_peak}.

Indeed, the previous p.d.f. is constituted of secondary electrons forming the peaks bases and the electron background. The production depth p.d.f. of the primary photo-electrons, ejected at the nanoparticle surface constituting most of the photo-electric peaks are created directly by photons shallower in the nanoparticle. That is why for the peak energies, their production depth p.d.f. are the same than the electron background p.d.f. until $r_p/R=0.5$ then diverge more and more till $r_p/R=1$.

\subsection{Comparison between our model and Livermore-Geant4 }
 
 From the spectra shown on Fig.\ref{Fig:spectrum} we see that our model results are consistent with the Livermore-Geant4 ones. But both spectra do not exactly match. The model continuous background is slightly below the Livermore-Geant4 one and the model photo-electric peak intensities are above the Livemore-Geant4 ones as seen on Tab.\ref{Tab:G4_Model_peak}.

These differences are explained by the photon absorption p.d.f.. We observe on Fig.\ref{Fig:Reaction_d.f.p.} that these p.d.f. are comparable but slightly different. As seen before, the Livermore-Geant4 curve follows a linear function with the relative distance $r_p/R$ to the nanoparticle center whereas the model one follows a non-linear quadratic  function. As a consequence the model absorption probability is larger than the Livermore-Geant4 one under $0.75 R$ and smaller over these values. 

However as we have seen on Fig.\ref{Fig:dpf_r_peak}, the surface nanoparticle (where model photon absorption p.d.f. is larger) is the production place of the photo-electric line electrons. So the gap between the model and the Livermore-Geant4 photo-electric line intensities presented on Tab.\ref{Tab:G4_Model_peak} is due to the difference between the photon absorption p.d.f. after $0.75R$. By the same way the gap between electron background is explained by the difference between photon absorption p.d.f. before $0.75R$. 

\begin{table*}
\centering
 \begin{tabular}{cccc}
 \hline
  Energies& Model  & Livermore-Geant4 & Ratios\\
  $eV$ & $electron.photon^{-1}.eV^{-1}$ &  $electron.photon^{-1}.eV^{-1}$ & \\
  \hline
    \hline
  $752$ & $6.34\times 10^{-5}$ & $7.26\times 10^{-5}$ & $1.15$\\
  $863$ & $8.26\times 10^{-5}$ & $9.40\times 10^{-5}$ & $1.14$ \\
  $962$ & $2.71\times 10^{-4}$ & $2.19\times 10^{-4}$ & $0.81$\\
  $1146$ & $3.94\times 10^{-4}$ & $3.06\times 10^{-4}$ & $0.78$\\
  $1164$ & $5.98\times 10^{-4}$ & $4.72\times 10^{-4}$ & $0.79$\\
  $1402$ & $3.18\times 10^{-4}$ & $3.25\times 10^{-4}$ & $1.02$\\
  $1406$ & $4.83\times 10^{-4}$ & $4.10\times 10^{-4}$ & $0.85$\\
  $1438$ & $6.02\times 10^{-5}$ & $4.98\times 10^{-5}$ & $0.83$\\
  $1489$ & $5.72\times 10^{-5}$ & $4.65\times 10^{-5}$ & $0.81$ \\
    \hline
\end{tabular} 
  \caption{Model and Livermore-Geant4 photo-electric line intensities from the baselines have been subtracted and their ratios for $5nm$ radius gold nanoparticle and $E_\gamma=1486.5eV$.}
  \label{Tab:G4_Model_peak}
 \end{table*}   

\subsection{Comparison between model nanoparticle and plane surface electron emission}

\begin{figure}
   \centering
   \includegraphics[width=0.4\textwidth]{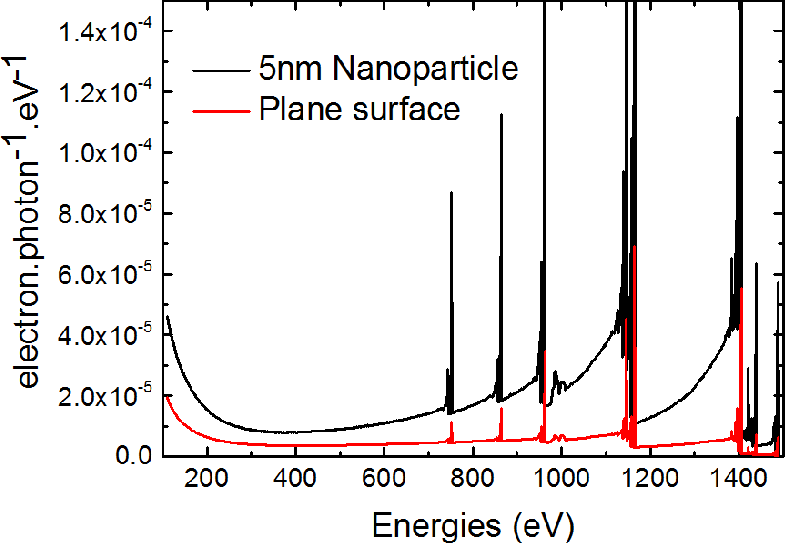}
     \caption{Electron emission spectra normalized to the absorbed photon number for a gold infinite plane surface and a gold nanoparticle of radius $5nm$ irradiated perpendicularly by $1486.5eV$ photons.}
  \label{Fig:reaction_spectra}
 \end{figure}

  To compare a gold nanoparticle of radius $5nm$ and a gold infinite plane surface electron emission, we normalize both spectra by the number of absorbed photons. The results are presented on Fig.\ref{Fig:reaction_spectra} for electron emission spectra and on Tab.\ref{Tab:r_e_Math} for photo-electric line intensities. 
  
We can see that the number of emitted electrons by photon absorption is much greater for the nanoparticle than for the gold plane surface. This is due to the small nanoparticle size.
Indeed each photon absorbed by the nanoparticle produces an electron cascade which as seen on Fig.\ref{Fig:Reaction_d.f.p.} is at a distance of a few nanometers of the nanoparticle surface. So there is a high probability that this cascade creates electrons at the nanoparticle surface. 

\begin{figure}
   \centering
   \includegraphics[width=0.4\textwidth]{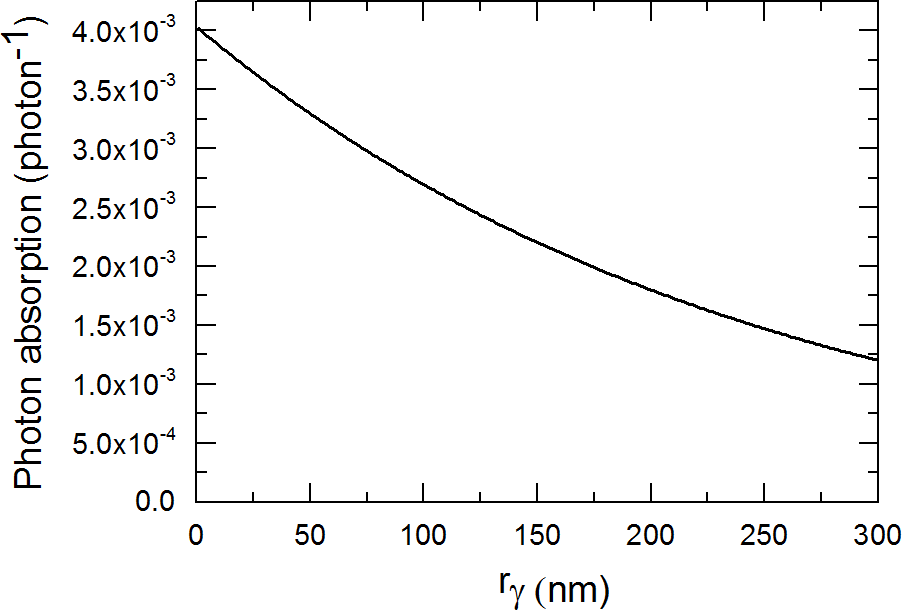}
     \caption{Photon absorption p.d.f. for a gold infinite plane surface as a function of $r_\gamma$ from our model.}
  \label{Fig:Reaction_Surface}
 \end{figure}

As seen in Fig.\ref{Fig:Reaction_Surface}, in the infinite plane surface, all the photons are absorbed but for most of them this happens too deep from the surface to create electron cascades which have a reasonable chance to reach the surface. Indeed by observing the electron cascade $\int \psi d \Omega$ p.d.f. on Fig.\ref{Fig:PsiCascade}, we see that after $30nm$ there is almost no chance for these electrons to reach the surface.\\

\begin{table*}
\centering
 \begin{tabular}{cccc}
 \hline
  Energies & 5nm Nanoparticle  & Plane Surface & Ratios\\
   $eV$  & $electron.photon^{-1}.eV^{-1}$ & $electron.photon^{-1}.eV^{-1}$ & \\
  \hline
    \hline
  $752$ & $7.26 \times 10^{-5}$ & $6.74 \times 10^{-6}$ & $10.77$\\
  $863$ & $9.40 \times 10^{-5}$ & $1.08 \times 10^{-5}$ & $8.73$ \\
  $962$ & $2.19 \times 10^{-4}$ & $3.00 \times 10^{-5}$ & $7.31$\\
  $1146$ & $3.06 \times 10^{-4}$ & $3.99 \times 10^{-5}$ & $7.66$\\
  $1164$ & $4.72 \times 10^{-4}$ & $6.59 \times 10^{-5}$ & $7.15$\\
  $1402$ & $3.25 \times 10^{-4}$ & $3.36 \times 10^{-5}$ & $9.66$\\
  $1406$ & $4.11 \times 10^{-4}$ & $5.44 \times 10^{-5}$ & $7.55$\\
  $1438$ & $4.98 \times 10^{-5}$ & $6.63 \times 10^{-5}$ & $7.52$\\
  $1489$ & $4.65 \times 10^{-5}$ & $6.32 \times 10^{-5}$ & $7.36$ \\
    \hline
\end{tabular} 
  \caption{Photo-electric line intensities from the baselines have been subtracted and their ratios, normalized to the number of absorbed photon, for a $5nm$ radius gold nanoparticle and an infinite gold plane surface with $E_\gamma=1486.5eV$.}
  \label{Tab:r_e_Math}
 \end{table*}   
 
\begin{figure}
   \centering
   \includegraphics[width=0.4\textwidth]{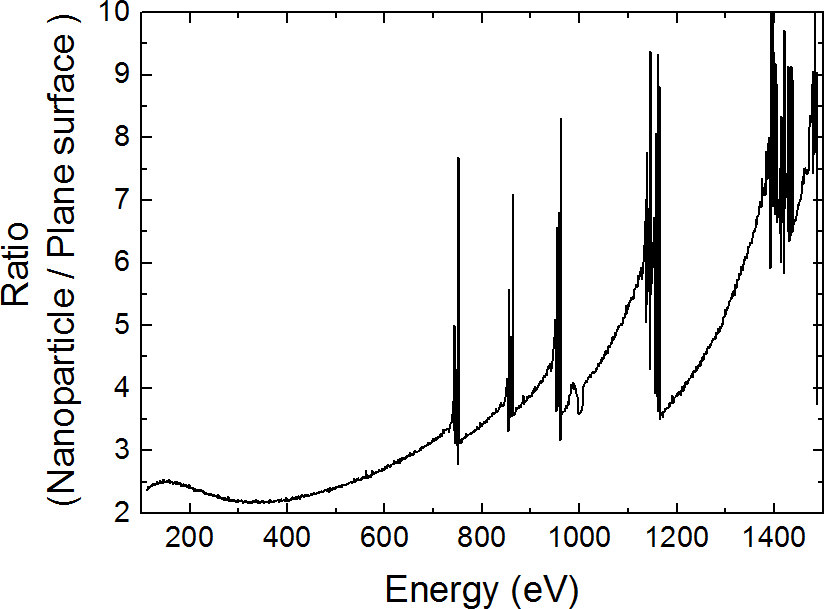}
     \caption{Ratio between gold nanoparticle of radius $5nm$ and infinite gold plane surface electron emission.}
  \label{Fig:Nano_Surface_Ratio}
 \end{figure}

In order to compare in details both normalized electron emission, we draw their ratios i.e. we divide the $5nm$ radius gold nanoparticle electron emission spectra intensity by the corresponding gold plane surface intensity. The resulting ratio is presented on Fig.\ref{Fig:Nano_Surface_Ratio} for spectra and on Tab.\ref{Tab:r_e_Math} for photo-electric line intensities.
 
We observe on Tab.\ref{Tab:r_e_Math} that the photo-electric line intensities normalized to the absorbed photon number are much larger (between $7$ and $10$ times) for nanoparticle than for the plane surface. We also observe on the spectra ratio Fig.\ref{Fig:Nano_Surface_Ratio} that after each peak the ratio slowly decreases, as a consequence of the electron cascade p.d.f. $\int \psi d \Omega$. 

\begin{figure}
   \centering
   \includegraphics[width=0.4\textwidth]{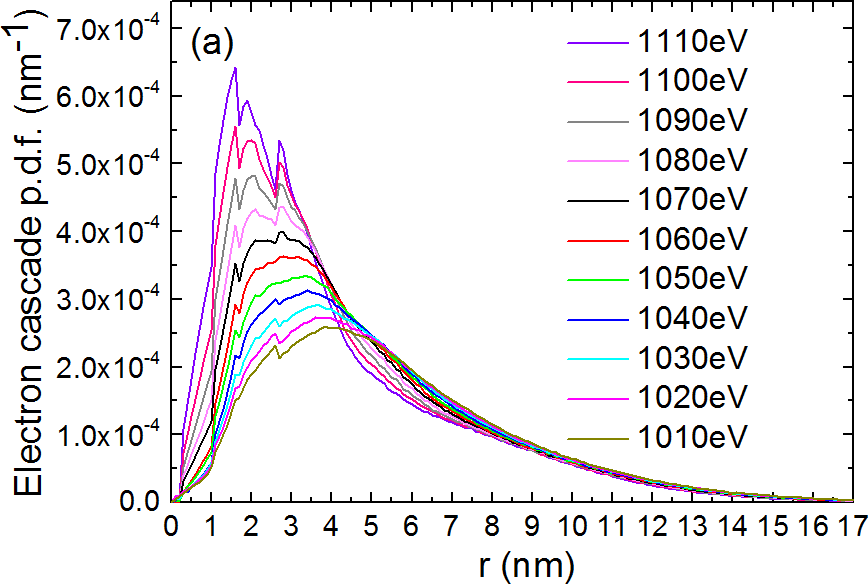}
   \includegraphics[width=0.4\textwidth]{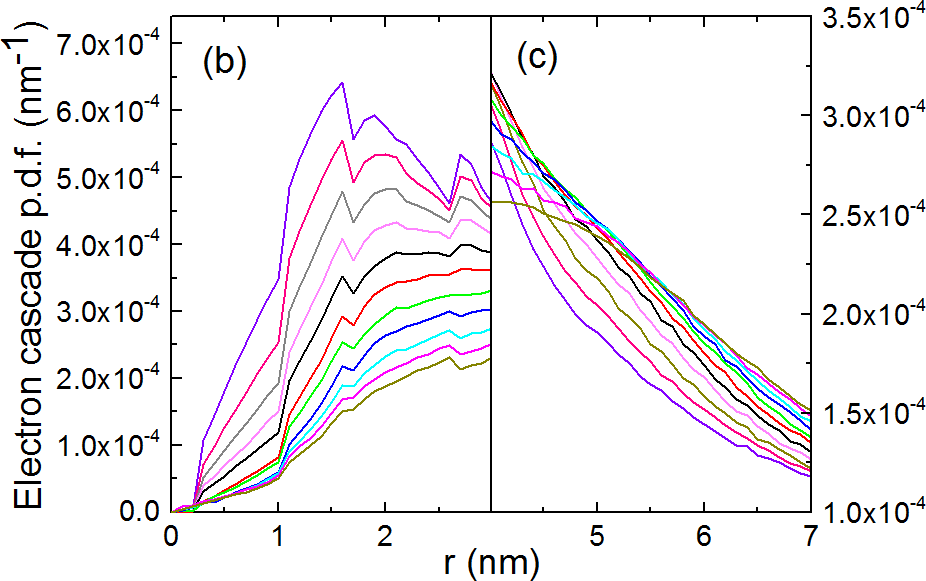}
     \caption{Electron cascade p.d.f. integrated over solid angle $\int \psi d\Omega$ for gold, $1010eV<E_e <1110eV$, and $E_\gamma=1486.5eV$ for range $0-17nm$ (a), $0-3nm$ (b) and $4-7nm$ (c).}
  \label{Fig:Electron_Cascade_peak}
 \end{figure}

We present this p.d.f. $\int \psi d \Omega$ as a function of the distance $r$ from absorption point, for energies between $1110eV$ and $1010eV$ on Fig.\ref{Fig:Electron_Cascade_peak}. We see that the further we move down from the peak energy ($1146eV$ in our case), the lower are the probabilities to find an electron at a small distance from the absorption point (Fig.\ref{Fig:Electron_Cascade_peak}(b)). On the other hand, as we move down from the peak energy we observe an increase of the $\int \psi d \Omega$ p.d.f. for $r$ values between $3.5nm$ and $7nm$ (Fig.\ref{Fig:Electron_Cascade_peak}(c)) and there is clearly a shift of $\int \psi d\Omega$ maximum to larger $r$ values (Fig.\ref{Fig:Electron_Cascade_peak}(a)).

In a $5nm$ radius nanoparticle, we know that most of the absorption points are close from the surface ($<3nm$). Consequently, the nanoparticle electron emission is strongly affected by the decrease of $\int \psi d\Omega$ p.d.f. at small $r$ values. The plane surface absorptions are located as seen previously further from the surface and are less affected by this decrease. 

For these reasons, as we move down from the peak energy, there is a stronger decrease for nanoparticle electron intensities than for plane surfaces, explaining the ratio decrease between both electron emissions.

\subsection{Radius impact on nanoparticle electron emission}

\begin{figure}
   \centering
   \includegraphics[width=0.4\textwidth]{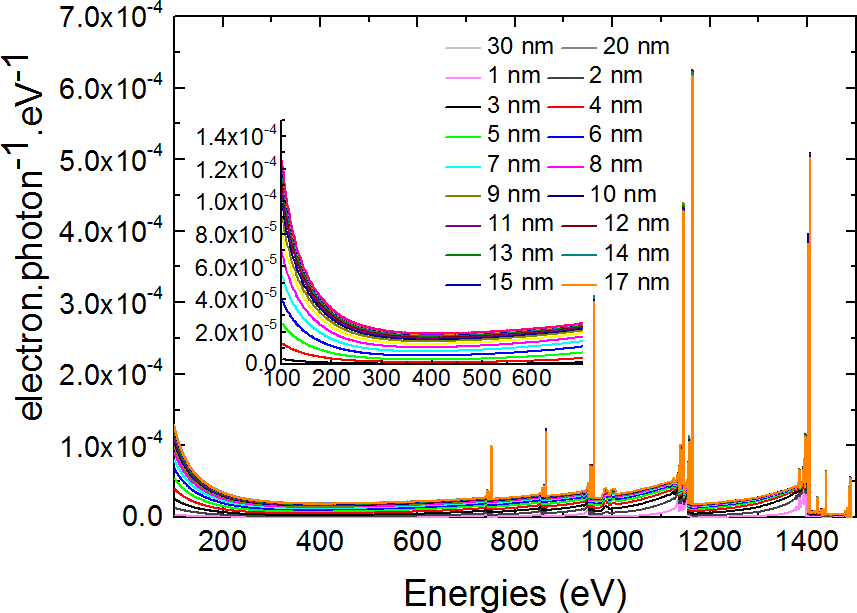}
     \caption{Gold nanoparticle electron emission spectra computed with the previous developed model for different nanoparticle radii.}
  \label{Fig:dpf_R}
 \end{figure}

One of the main interest of our model is that we can keep the same electron cascade p.d.f. but easily change the nanoparticle radius, which is a critical parameter. This  is an important gain of computing time compared to a full Monte-Carlo simulation approach classically used. To study the radius influence over the electron emission, we compute it for several radii from $1nm$ to $100nm$ and present few of them on Fig.\ref{Fig:dpf_R}. We observe that the electron emission rapidly increases from $1nm$ to $15nm$ and then saturates. The spectrum form does not seem to depend of the nanoparticle radius.\\

\begin{figure}
   \centering
   \includegraphics[width=0.4\textwidth]{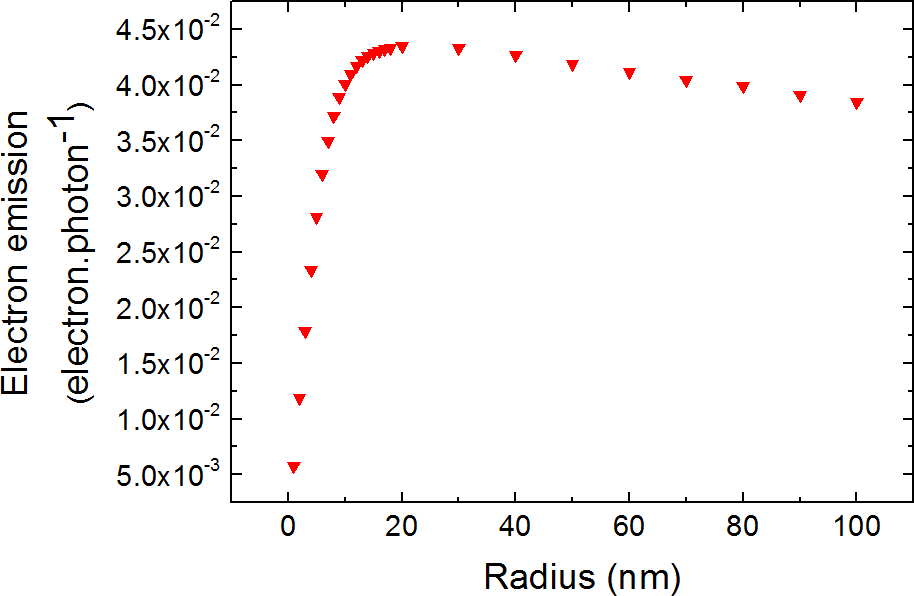}
     \caption{Total gold nanoparticle electron emission computed with the developed model for different nanoparticle radii.}
  \label{Fig:EE_R}
 \end{figure}

To have a more accurate information about the radius influence, we draw the nanoparticle electron emission integrated over the ejected electrons energy $E_e$ from $100eV$ to $1500eV$ as a function of the nanoparticle radius on Fig.\ref{Fig:EE_R}. We see, as expected, an important increase from $1nm$ to $15nm$ and then a slow decrease.

\begin{figure}
   \centering
   \includegraphics[width=0.4\textwidth]{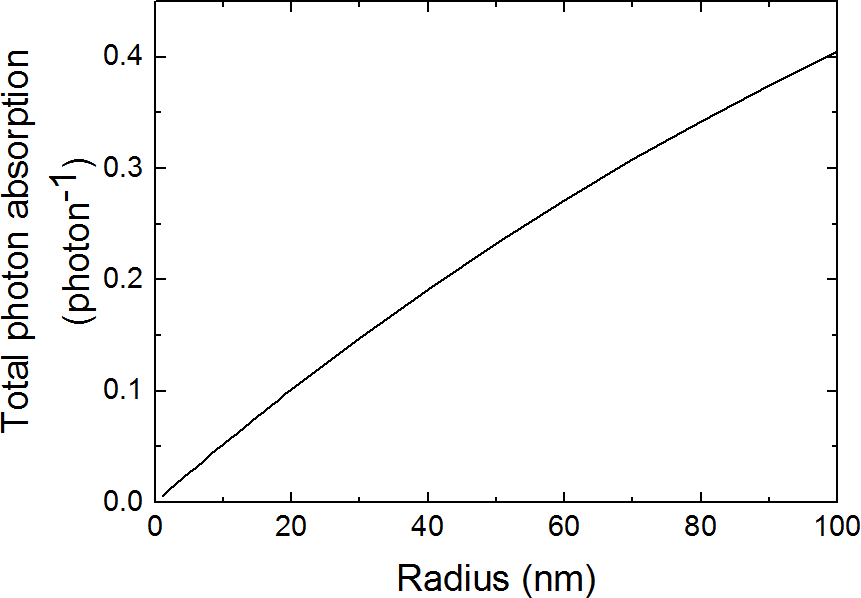}
     \caption{Total gold nanoparticle photon absorption computed with the developed model for different nanoparticle radii.}
  \label{Fig:Total_photon_reaction}
 \end{figure}

\begin{figure}
   \centering
   \includegraphics[width=0.4\textwidth]{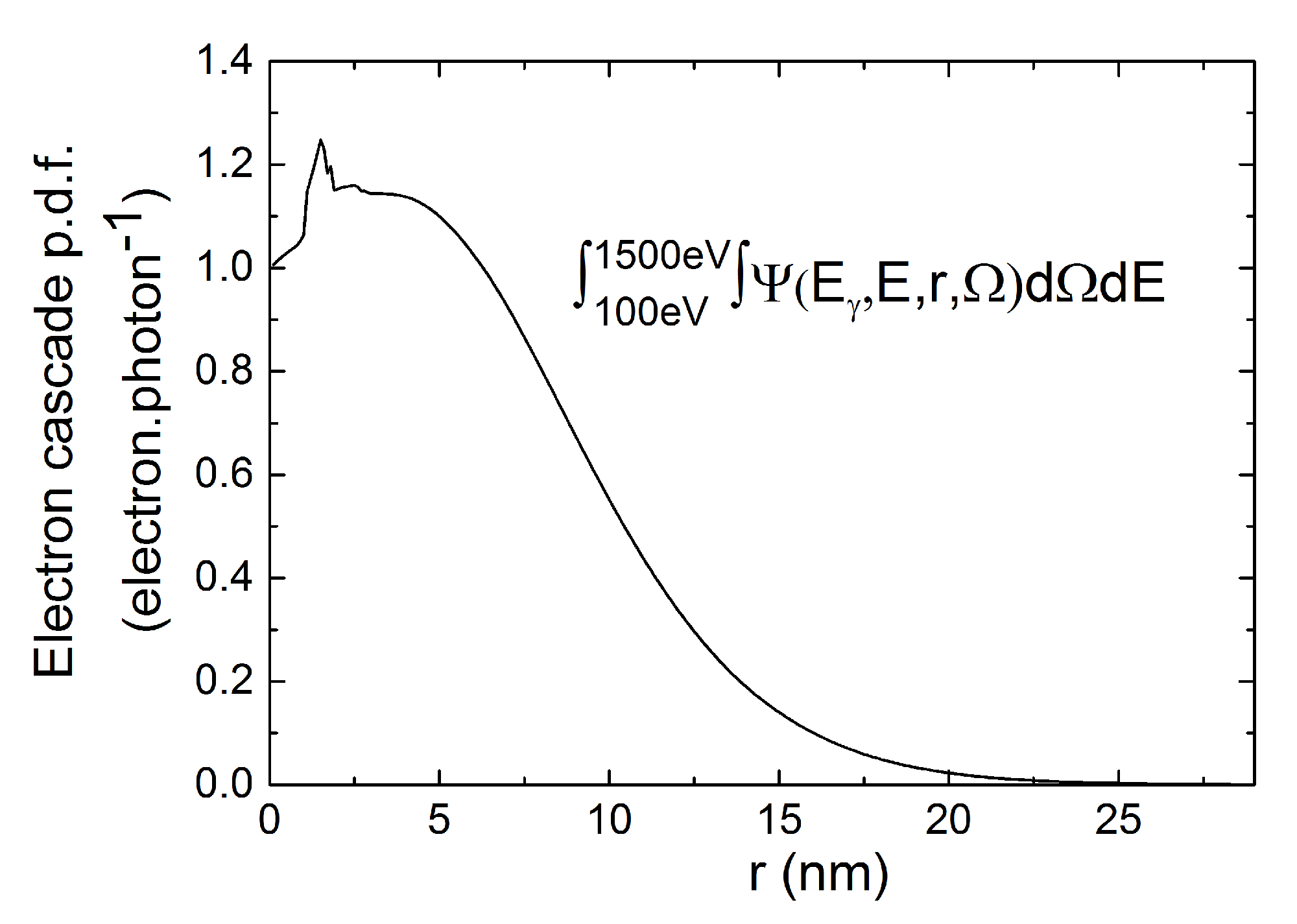}
     \caption{Gold electron cascade p.d.f. integrated over electron energy $E$ from $100eV$ to $1500eV$ and solid  angle of collection $\Omega$ .}
  \label{Fig:Electron_cascade_dpf}
 \end{figure}

 These features can be explained qualitatively by the total photon absorption p.d.f. as a function of nanoparticle radius, shown on Fig.\ref{Fig:Total_photon_reaction}, and by the gold electron cascade p.d.f. 
\begin{center}
 $\int_{100eV}^{1500eV} \int  \psi(E_\gamma,E,r,\Omega) d\Omega dE$  (Fig.\ref{Fig:Electron_cascade_dpf})
 \end{center}
  i.e. the p.d.f. that an electron cascade produces an electron at a distance $r$ of its production place.

The total photon absorption (Fig.\ref{Fig:Total_photon_reaction}) shows an increase with the nanoparticle radius, which in turn causes an increase in the first part of the total electron emission (Fig.\ref{Fig:EE_R}). Indeed if there are more photon absorbed, there are more photo-electrons produced, more electron cascades and consequently more electrons ejected at the surface.\\

This increase of the total electron emission stops at a maximum located at $R=20nm$. Further a slow decrease appears because the number of electron cascades able to reach the surface is less and less important. As seen on Fig.\ref{Fig:Electron_cascade_dpf}, once the distance from the absorption point is larger than $6nm$ the further an electron cascade is created from the absorption point, the lower its probability to reach the surface and eject electrons.

Up to $6nm$ as long as the electron emission increases, the total electron emission  curve of Fig.\ref{Fig:EE_R} follows the total absorption increase, because it is not counterbalanced by the decrease of the electron cascade p.d.f.. In small nanoparticles most of the electron cascade production points are closer than $6nm$ from the surface.  As a consequence their probability to reach the surface is roughly stable as shown by  Fig.\ref{Fig:Electron_cascade_dpf}. For larger nanoparticles, the number of electron production points further than $6nm$ from the surface becomes significant and the absorption increase is counterbalanced by the decrease of the electron cascade p.d.f..

\section{Conclusion}

We developed an original model for the electron emission of high-Z nanoparticle and plane surface irradiated by X-ray photons. After checking that this model is compatible with the Livermore model implemented in Geant4, we used it to study the electron emission of a gold nanoparticle irradiated by $1486.5eV$ photons.
This model allowed us to deeply understand key features and parameters of nanoparticle electron emission: nanoparticle size, difference between nanoparticle and plane surface electron emission, electron cascade production depth, incident photon energy. 
This work highlights the existence of a nanoparticle radius corresponding to a maximum electron emission. This model can be simplified by doing approximations and other electron cascade models can be included. It can be checked with other simulation codes, photon energies or nanoparticle compositions.

\bibliographystyle{spbasic}

\end{document}